\documentclass[graybox]{svmult}

\usepackage{type1cm}        
\usepackage{makeidx}         
\usepackage{graphicx}        
\usepackage{multicol}        
\usepackage[bottom]{footmisc}

\usepackage[numbers,sort&compress,comma,square]{natbib}

\usepackage{newtxtext}       %
\usepackage{newtxmath}       

\usepackage{syntax}
\usepackage{pgf}
\usepackage{tikz}
\usetikzlibrary{positioning,arrows.meta,calc,patterns}
\usepackage{subcaption}
\usepackage{rotating}

\usepackage{multirow}
\usepackage{tablefootnote}

\usepackage[most]{tcolorbox}
\usepackage{xcolor}

\definecolor{ChapBoxDark}{RGB}{155,185,220}  
\definecolor{ChapBoxLight}{RGB}{235,245,255} 

\newtcolorbox{chapterpointer}[1]{%
  colback=ChapBoxLight,
  colframe=ChapBoxDark,
  coltitle=black,
  title=#1,
  fonttitle=\sffamily\bfseries,
  boxrule=0.5pt,
  arc=2mm,
  left=6pt,
  right=6pt,
  top=6pt,
  bottom=6pt
}

\makeindex             

\begin{document}

\title*{Artificial Intelligence and Modeling \& Simulation: An Overview}
\titlerunning{AI and M\&S: An Overview}

\author{Niclas Feldkamp, Philippe J. Giabbanelli, Istvan David}

\institute{Niclas Feldkamp \at Technische Universität Ilmenau, Germany, \email{niclas.feldkamp@tu-ilmenau.de}
\and Philippe J. Giabbanelli \at VMASC, Old Dominion University, USA, \email{pgiabban@odu.edu}
\and Istvan David \at McSCert, McMaster University, Canada, \email{istvan.david@mcmaster.ca}}


%
%
\maketitle

\label{chap:intro}

\abstract{Artificial intelligence (AI) and Modeling \& Simulation (M\&S) are increasingly intertwined, reflecting converging research needs across both communities, rapid technological advances such as the rise of generative AI, and the growing availability of data and computational resources. This report provides a structured overview of the intersections of AI and M\&S. The relationship goes both ways: AI can support, augment, or even replace components of simulation studies, while simulations can serve as data generators, training environments, and evaluation platforms for AI. We organize this landscape along the stages of M\&S from model specification and input modeling to execution, experimentation, verification and validation, and output analysis. Selected studies at each stage illustrates how techniques such as Large Language Models have reshaped simulation practices, while highlighting limitations and open challenges.  This report also provides a conceptual roadmap that helps readers navigate a rapidly changing ecosystem.}

\section{Primer: What is AI and why should we combine it with modeling \& simulation?}
AI is undoubtedly impacting just about every facet of our society and this is poised to continue. While AI and the underlying algorithms have been around for a very long time, the immense computing power\footnote{There are increasing concerns about the resource consumption of AI (e.g., energy to power and cool AI centers, which in turn use clean water) and its impact (e.g., greenhouse gases, transformation of agricultural land into data centers)~\cite{sinistore2025sustainable,kseibati2025cooling}. Reducing the footprint through green computing has long been a concern of the modeling \& simulation community. Despite a shared appreciation of the problem, there is a diversity of perspectives about potential solutions and the role that modeling can play~\cite{bork2024role}. One approach is to promote the reuse of simulation results (e.g., through clear provenance and metadata) instead of wastefully re-computing them~\cite{villamar2025metadata}. Wilsdorf and colleagues have shown how to reuse or even adapt simulation experiments~\cite{wilsdorf2023automatic}. Another approach is to realize that we may have performed enough simulations and that, sometimes through the use of AI, we can answer the question of interest without consuming more resources~\cite{lutz2022we}.} and access to data\footnote{AI raises several concerns about the use of material without clear consent (e.g., copyrighted, personally identifiable) and the opacity about the data that was used for training~\cite{pasetti2025technical,chesterman2025good}.} that are now available have enabled significant changes in scale and capabilities. From a purely definitional point of view, AI can be construed as an umbrella term for all artificial, computational systems that are designed to perform tasks that require a form of human-like intelligence. Although this broad definition encompasses aspects such as knowledge representation and reasoning (e.g., formal logic), the recent interest has been on the subspace of AI concerned with machine learning.\footnote{Articles often use terms such as AI, machine learning, or deep learning interchangeably, without clear boundaries. Besides, which term is used as umbrella changes over time. Recent bibliographic analyses point to (generative) AI being used as the current umbrella term~\cite{obreja2025mapping}. We consider Artificial Intelligence to be the overall field, of which machine learning is a subdomain, within which neural networks are found, and that can specifically be used through deep learning for Generative AI.}

Machine learning algorithms are usually divided into three basic categories: \textit{Supervised Learning}, \textit{Unsupervised Learning}, and \textit{Reinforcement Learning}~\cite{bishop2006pattern}. Supervised learning essentially involves prediction tasks that are mapped either as classification or regression. The learning process is based on sample data that contain a target variable to be predicted (i.e., labeled data). A prediction can then be made for new, unknown input data. In contrast, unsupervised learning does not have an explicit target variable. Instead, the algorithm attempts to independently identify structures and patterns in the underlying training data without specifying labels (e.g., to find clusters or associations). Supervised and unsupervised learning can also be combined very well. Unsupervised learning is first used to create structures or patterns in the input data, which can then be used as labels for the training of a supervised learning algorithm. The generic term for generating the training labels from the data itself is also referred to as \textit{self-supervised learning}, which can therefore be seen as an intermediate form between unsupervised and supervised learning. Another approach within self-supervised-learning is to leave out parts of the original data, which then have to be predicted accordingly, for example by leaving out parts of a picture or a leaving out a word from a sentence. This is also the method by which Large Language Models (LLMs) are trained. Through training in completing sentences, they can then autoregressively create new texts by predicting the next word (or token), based on the input prompt and the previously generated text~\cite{brown2020language}. Self-supervised training methods are very powerful, as unlabeled data is usually available in very large quantities, while manually labeling lots of data is obviously time-consuming and costly. Reinforcement learning takes a different approach: here, a sequence of actions is optimized by an agent that learns through interaction with an environment. The learning process is based on feedback in the form of rewards or punishments that the agent receives depending on the actions it performs~\cite{SuttonBarto2018ReinforcementLearning}.

Another perspective for classifying machine learning methods is based on the type of problem to be solved: \textit{descriptive} methods are used to describe and structure data, \textit{predictive methods} are used to predict unknown values, be it numerical (regression) or categorical (classification), and \textit{prescriptive} methods focus on decision-making by learning policies and predicting actions~\cite{wissuchek2025prescriptive}. As stated before, machine learning can actually be seen as a subcategory of a broader AI-definition.$^3$ In contrast to other subfields of AI such as expert systems which are fundamentally rule-based, machine learning is data-driven: the goal is to extract patterns, correlations, and rules from the training data in order to apply them to new, unknown data.\footnote{A frequent confusion and recurring debate is on whether machine learning \textit{extrapolates}. This term may be used intuitively to say that the algorithms can go `beyond' the data by handling cases that were not seen previously. However, from a mathematical viewpoint, that does not necessarily mean an extrapolation. Many learning models provide estimate \textit{within} the support of the training distribution~\cite{xu2012robustness} (i.e., interpolation). In the context of Generative AI, the debate on extrapolation vs. interpolation is similar to the challenge of generalization vs. memorization, as researchers observed that outputs seem to reflect interpolation of training samples rather than a generalization beyond that set~\cite{carlini2021extracting}. Fundamentally, extrapolation is a harder task than interpolation, and results can quickly worsen when going too far from the training data and in the absence of additional guidance such as causal frameworks.} This is particularly true with the advent of deep learning, or rather, the availability of computer infrastructures that enable deep learning. Large, multi-layered neural networks with many hidden layers allow the processing of very large amounts of data. The resulting ML models are extremely powerful, but on the other hand also so complex that their internal relations and decision-making mechanisms are essentially complete black boxes and can no longer be comprehended by humans. As a result, a new scientific discipline has emerged: \textit{Explainable AI} (XAI), which aims to develop methods and algorithms that make the decisions of black-box machine learning models transparent and explainable again~\cite{arrieta2020explainable}. Forecasts and decisions made by ML models can have real and sometimes drastic consequences. This responsibility should always be kept in mind when developing such models, as well as possible regulatory requirements on the right to explanation~\cite{goodman2017european}.

As stated above, the hype surrounding AI is currently focusing on the topic of generative AI. However, generative algorithms are not an entirely new development. Assume that we have input data $X$ and want to predict the target $Y$. A traditional, non-generative prediction algorithm (known as a \textit{discriminative model}) learns the decision boundary within the input data $X$ during training, i.e., the probability $P(Y|X)$~\cite{ng2001discriminative}. Let's imagine a fictional neural network that can classify images of dogs and cats. For each animal image $x \in X$, it gives us probabilities for the cat class and the dog class, which add up to 100\%. The more confident the network is in its decision, the more the probability tilts toward one class. The network has thus learned to draw a line between dogs and cats within the input data $X$, i.e., to discriminate between them. However, this does not necessarily mean that the network has learned the specific characteristics of dogs and cats, let alone that it could independently draw a new dog or cat. This is where we find the crucial difference to generative models. During training, generative models learn the underlying probability distribution of the training data $X$. In other words, they learn $P(X|Y)$. This can be rearranged for inference to derive $P(Y|X)$, but a particularly powerful application is that generative models can also generate new samples from this context~\cite{ng2001discriminative}. Such generative models have been around for a very long time: classic Bayesian networks, for example, also count as generative models because they model the joint probability distribution $P(X,Y)$ and can therefore, in principle, both predict and generate new samples. The current attention devoted to generative AI relates to the breakthrough of extremely powerful algorithms, all of which are based on deep learning. In addition, these are usually very large-scale models that not only require enormous amounts of data for training, but also very powerful chips and corresponding computer infrastructures$^1$.

In the public perception, it seems that more or less everything in the context of generative AI revolves around LLMs. However, generative AI actually includes much more than that, and the current state-of-the-art algorithms are quite diverse: the generation of text, audio, images, and even three-dimensional scenes can be handled by a wide variety of algorithms, each with specific characteristics, but all based on deep learning in principle. For example, \textit{Generative adversarial networks} (GANs) and \textit{variational autoencoders} (VAEs) have long achieved very good results in image generation, while diffusion models have gained popularity.\footnote{Although newer studies may mention \textit{diffusion models} more often than GANs or VAEs, it does not mean that GANs and VAEs have been \textit{replaced}. GANs and VAEs are still actively researched, widely used, and in some cases preferable depending on the task. They also remain components in many modern pipelines: for example, VAE encoders/decoders are common in diffusion-based generators.} Diffusion models are neural networks that are trained on the approach that noise is gradually added to an image, and the model must then remove this noise to restore the previous state (de-noising). In this way, it learns the probability distribution of this denoising process and can thus generate plausible new images from complete noise~\cite{Song2021ScoreBasedSDE}. Diffusion models and VAEs have been used in modeling and simulation research to generate illustrations of a modeling report on-the-fly~\cite{gandee2024visual} or to create photorealistic avatars for simulated agents~\cite{giabbanelli2025emerging}. \textit{Neural radiance fields} (NeRFs) go even further in that they are trained to generate three-dimensional objects or scenes from two-dimensional images~\cite{mildenhall2021nerf}. 

Besides generative AI, discriminative methods certainly continue to play a central role and are widely used in practice. Interestingly, these methods do not necessarily have to be based on neural networks or deep learning approaches. Methods such as random forests or support vector machines continue to demonstrate very high performance and efficiency in certain applications for regression and classification tasks and are sometimes equal or superior to modern deep learning methods. Quick and transparent methods such as decision trees can also offer good results in problems that are linearly separable, while making it easy for users to interpret the results (i.e., \textit{white box}). In the area of unsupervised learning, classic clustering algorithms such as k-means are still very effective, although deep learning-based methods such as autoencoders (for example for anomaly detection) are also used for more complex data~\cite{an2015variational}. Neural network architectures remain valuable to process sequential data such as time series, text, or audio. While recurrent neural networks and long short-term memory networks were previously mainly used for this purpose, these have now largely been replaced by transformer architectures. Those are very well suited for long sequences and large amounts of data and are the enabling technology for large language models~\cite{vaswani2017attention}. Table~\ref{table:overview} shows a selection of common methods with a variety of learning paradigms and application possibilities. 

\begin{table}[htb]
\caption{Overview of different AI and machine learning methods divided into different learning paradigms and other features. \footnotesize\textit{$^*$ Training for LLMs usually includes multiple steps, including self-supervised, supervised, and reinforcement learning.}}
\begin{tabular}{|p{2.2cm}|p{2cm}|p{2.2cm}|l|p{1.25cm}|p{2cm}|}
\hline
\textbf{Algorithm}                                         & \textbf{Learning paradigm}     & \textbf{Problem-solving capability} & \textbf{Model type}             & \textbf{Deep learning} & \textbf{Typical tasks}                                    \\ \hline
Decision   Trees                                           & \multirow{3}{*}{Supervised}    & Descriptive / Predictive            & \multirow{4}{*}{Discriminative} & \multirow{2}{*}{No}    & (White-box)   classification                              \\ \cline{1-1} \cline{3-3} \cline{6-6} 
Random   Forests                                           &                                & \multirow{2}{*}{Predictive}         &                                 &                        & Robust   regression                                       \\ \cline{1-1} \cline{5-6} 
Multilayer   Perceptron (MLP) / Feedforward Neural Network &                                &                                     &                                 & Yes                    & Prediction   for complex data (e.g. image classification) \\ \cline{1-3} \cline{5-6} 
k-Means                                                    & \multirow{3}{*}{Unsupervised}  & \multirow{2}{*}{Descriptive}        &                                 & \multirow{2}{*}{No}    & Clustering                                                \\ \cline{1-1} \cline{4-4} \cline{6-6} 
Gaussian   Mixture Models (GMM)                            &                                &                                     & Generative                      &                        & Probabilistic   clustering                                \\ \cline{1-1} \cline{3-6} 
Variational   Autoencoder                                  &                                & Descriptive / Predictive            & Generative                      & \multirow{4}{*}{Yes}   & Anomaly detection  /   General purpose data generation    \\ \cline{1-4} \cline{6-6} 
Diffusion   Models                                         & Unsupervised                   & \multirow{2}{*}{Predictive}         & Generative                      &                        & Image   generation                                        \\ \cline{1-2} \cline{4-4} \cline{6-6} 
Long   short-term memory nets                              & Supervised                     &                                     & Discriminative                  &                        & Time   series Processing                                  \\ \cline{1-4} \cline{6-6} 
NeRF                                                       & Self-Supervised                & Descriptive                         & Generative                      &                        & 3D   generation                                           \\ \hline
Q-Learning                                                 & \multirow{3}{*}{Reinforcement} & \multirow{3}{*}{Prescriptive}       & Discriminative                  & No                     & Decision-making                                           \\ \cline{1-1} \cline{4-6} 
Deep   Q-Networks (DQN)                                    &                                &                                     & Discriminative                  & Yes                    & (Complex)   decision-making                               \\ \cline{1-1} \cline{4-6} 
Policy   Gradient Optimization                             &                                &                                     & Generative                      & No                     & Stochastic   policies                                     \\ \hline
Transformer   (Encoder-Only)                               & \multirow{2}{*}{Supervised}    & \multirow{3}{*}{Predictive}         & Discriminative                  & \multirow{3}{*}{Yes}   & Text   classification                                     \\ \cline{1-1} \cline{4-4} \cline{6-6} 
Transformer   (Encoder-Decoder)                            &                                &                                     & Generative                      &                        & Translation                                               \\ \cline{1-2} \cline{4-4} \cline{6-6} 
Large   Language Models (LLMs)                             & All$^*$                           &                                     & Generative                      &                        & Text and code generation                                  \\ \hline
\end{tabular}
\label{table:overview}
\end{table}

In summary, data-driven algorithms (i.e., algorithms that can learn patterns, correlations, and rules from data) are continuously being improved (mainly under the label AI) and continue to maintain widespread popularity. Current and future developments such as generative AI are also fueling this trend. Simply said, these algorithms need to consume data in order to function properly. Since models generate data through simulations as part of answering a question, this generated data can be an asset for AI training. Both worlds – AI on the one hand, and modeling and simulation (M\&S) on the other – thus have natural fit~\cite{feldkamp2024application}. The possible applications resulting from the combination of both disciplines are extremely diverse. The following sections provide examples of how AI and M\&S can be combined in a useful manner, both within simulation studies in the traditional sense and outside of simulation studies. The latter means that not only M\&S can benefit from the combination with AI, but also vice versa. For this purpose, we show examples in which AI is dependent on simulation models or at least benefits greatly from the combination with simulation.

\section{AI to support modeling and simulation studies}
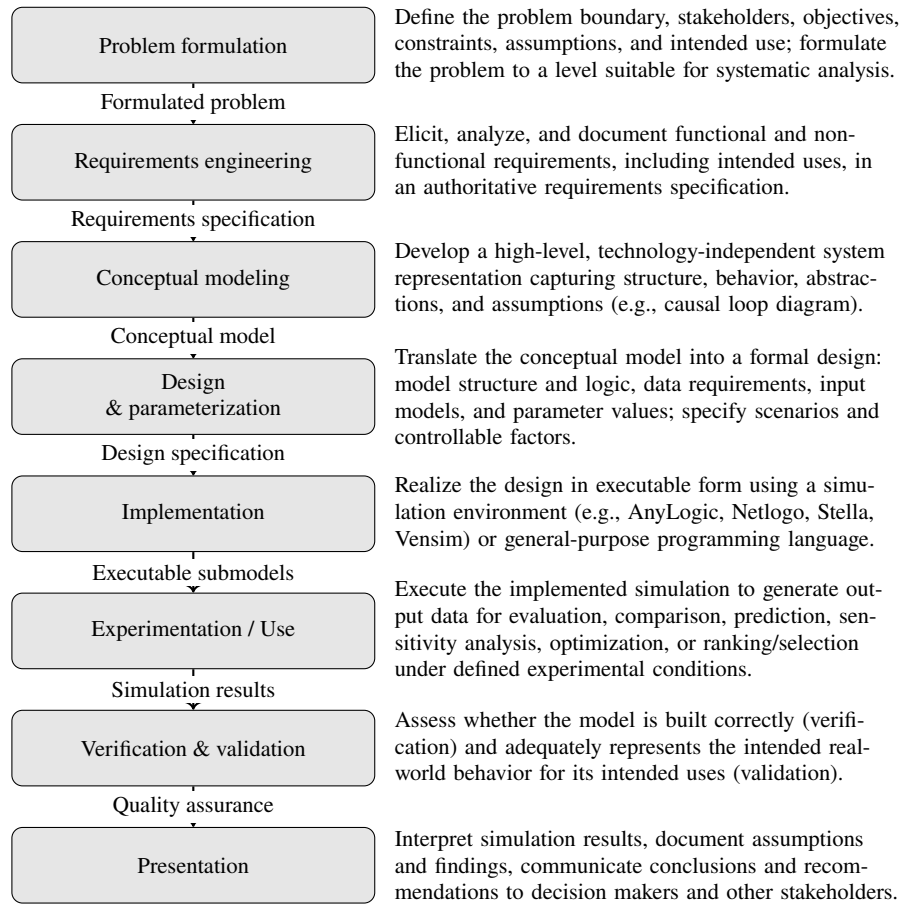
\begin{figure}[t]
\centering
\begin{tikzpicture}[
  node distance=1.55cm,
  box/.style={
    draw,
    rectangle,
    rounded corners,
    align=center,
    minimum width=4.8cm,
    minimum height=1.0cm
  },
  det/.style={box, fill=gray!20},
  def/.style={align=left, text width=6.7cm, anchor=west},
  arrow/.style={->, thick},
  lab/.style={midway, fill=white, inner sep=2pt, font=\small}
]

\node[det] (pf)   {Problem formulation};
\node[det] (re)   [below of=pf] {Requirements engineering};
\node[det] (cm)   [below of=re] {Conceptual modeling};
\node[det] (des)  [below of=cm] {Design\\\& parameterization};
\node[det] (impl) [below of=des] {Implementation};
\node[det] (exp)  [below of=impl] {Experimentation / Use};
\node[det]  (vv)   [below of=exp] {Verification \& validation};
\node[det]  (pres) [below of=vv] {Presentation};

\node[def] at ($(pf.east)+(0.15cm,0)$) {
Define the problem boundary, stakeholders, objectives, constraints, assumptions, and intended use; formulate the problem to a level suitable for systematic analysis.
};

\node[def] at ($(re.east)+(0.15cm,0)$) {
Elicit, analyze, and document functional and non-functional requirements, including intended uses, in an authoritative requirements specification.
};

\node[def] at ($(cm.east)+(0.15cm,0)$) {
Develop a high-level, technology-independent system representation capturing structure, behavior, abstractions, and assumptions (e.g., causal loop diagram).
};

\node[def] at ($(des.east)+(0.15cm,0)$) {
Translate the conceptual model into a formal design: model structure and logic, data requirements, input models, and parameter values; specify scenarios and controllable factors.
};

\node[def] at ($(impl.east)+(0.15cm,0)$) {
Realize the design in executable form using a simulation environment (e.g., AnyLogic, Netlogo, Stella, Vensim) or general-purpose programming language.
};

\node[def] at ($(exp.east)+(0.15cm,0)$) {
Execute the implemented simulation to generate output data for evaluation, comparison, prediction, sensitivity analysis, optimization, or ranking/selection under defined experimental conditions.
};

\node[def] at ($(vv.east)+(0.15cm,0)$) {
Assess whether the model is built correctly (verification) and adequately represents the intended real-world behavior for its intended uses (validation).
};

\node[def] at ($(pres.east)+(0.15cm,0)$) {
Interpret simulation results, document assumptions and findings, communicate conclusions and recommendations to decision makers and other stakeholders.
};

\draw[arrow] (pf)   -- node[lab] {Formulated problem} (re);
\draw[arrow] (re)   -- node[lab] {Requirements specification} (cm);
\draw[arrow] (cm)   -- node[lab] {Conceptual model} (des);
\draw[arrow] (des)  -- node[lab] {Design specification} (impl);
\draw[arrow] (impl) -- node[lab] {Executable submodels} (exp);
\draw[arrow] (exp)  -- node[lab] {Simulation results} (vv);
\draw[arrow] (vv)   -- node[lab] {Quality assurance} (pres);

\end{tikzpicture}
\caption{Pipeline of a modeling and simulation study. We omit transitions that go back in the system, e.g. verification and validation issues lead to changes in prior stages. Similar perspectives on the M\&S pipeline have been exposed on many occasions. Related views such as Sargent's also include \textit{system theories} (which are modeled into a conceptual model and hypothesized when analyzing simulations)~\cite{Sargent01022013}, while Balci points out that verification and validation is an activity across all stages rather than its own stage (e.g., a conceptual model can also be verified and validated)~\cite{balci2012life}.}
\label{fig:sim-study-pipeline}
\end{figure}

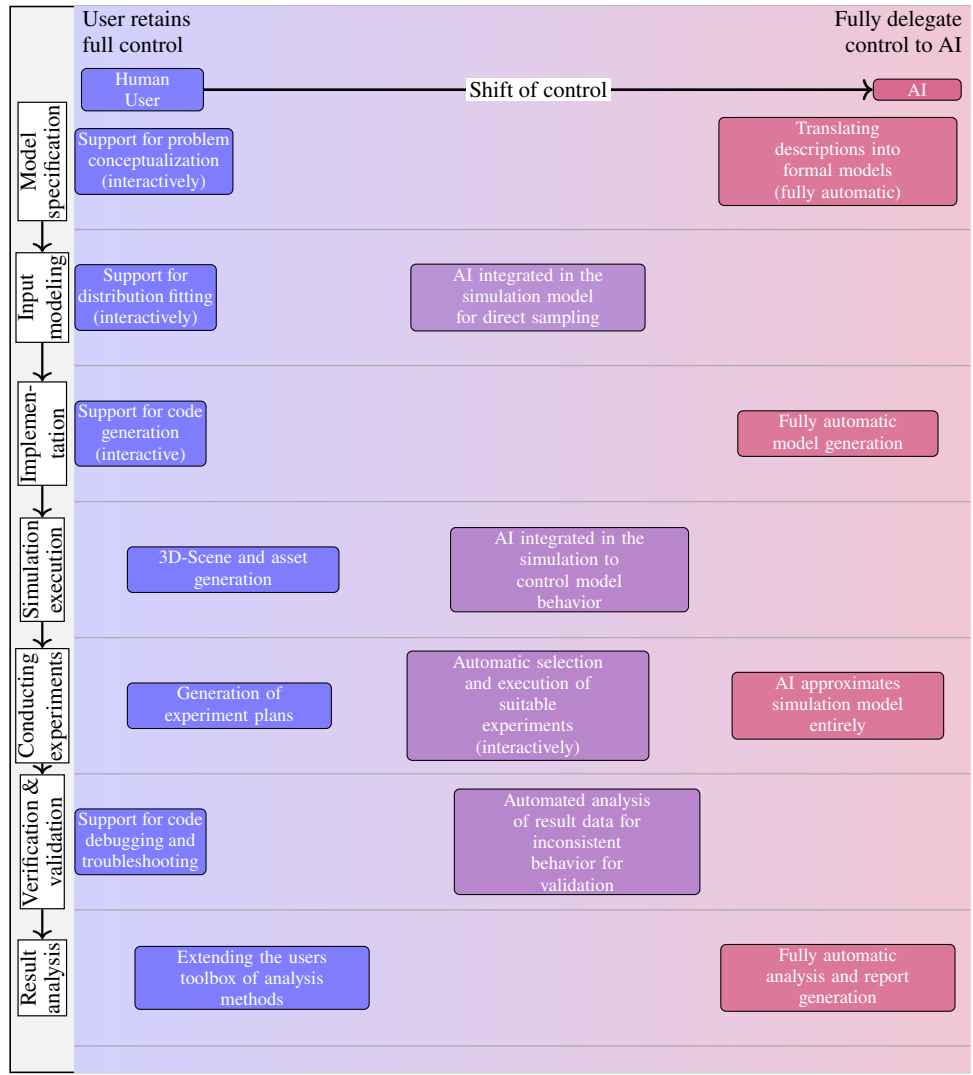
\begin{figure}[t]
\centering
\begin{tikzpicture}

\def\Xleft{0.00}
\def\Xright{12.70}
\def\Ytop{2.05}
\def\Ybot{-12.05}

\def\WboxW{0.85}
\def\Wcx{0.425} 

\def\Gleft{\WboxW}
\def\Gright{\Xright}

\def\HdrLabelY{1.72}
\def\HdrBoxY{0.95}

\def\xA{\Gleft}
\def\xM{6.85}
\def\xR{10.95}

\def\Rone{0.00}
\def\Rtwo{-1.80}
\def\Rthree{-3.60}
\def\Rfour{-5.40}
\def\Rfive{-7.20}
\def\Rsix{-9.00}
\def\Rseven{-10.80}

\def\Done{-0.90}
\def\Dtwo{-2.70}
\def\Dthree{-4.50}
\def\Dfour{-6.30}
\def\Dfive{-8.10}
\def\Dsix{-9.90}
\def\Dseven{-11.70}

\fill[gray!10] (\Xleft,\Ytop) rectangle (\Xright,\Ybot);
\draw[line width=0.6pt] (\Xleft,\Ytop) rectangle (\Xright,\Ybot);

\shade[left color=blue!18, right color=purple!22]
  (\Gleft,\Ytop) rectangle (\Gright,\Ybot);

\draw[gray!65, line width=0.5pt] (\Gleft,\Done)   -- (\Gright,\Done);
\draw[gray!65, line width=0.5pt] (\Gleft,\Dtwo)   -- (\Gright,\Dtwo);
\draw[gray!65, line width=0.5pt] (\Gleft,\Dthree) -- (\Gright,\Dthree);
\draw[gray!65, line width=0.5pt] (\Gleft,\Dfour)  -- (\Gright,\Dfour);
\draw[gray!65, line width=0.5pt] (\Gleft,\Dfive)  -- (\Gright,\Dfive);
\draw[gray!65, line width=0.5pt] (\Gleft,\Dsix)   -- (\Gright,\Dsix);
\draw[gray!65, line width=0.5pt] (\Gleft,\Dseven) -- (\Gright,\Dseven);

\node[align=left, font=\small, anchor=west] at (\Gleft,\HdrLabelY) {User retains\\full control};
\node[align=right, font=\small, anchor=east] at (\Gright,\HdrLabelY) {Fully delegate\\control to AI};

\node[
  draw=black, rounded corners=2pt,
  fill=blue!55, fill opacity=0.85,
  text=white, text opacity=1,
  align=center, font=\scriptsize,
  inner sep=1.6pt, text width=1.50cm
] (HU) at (\Gleft+0.90,\HdrBoxY) {Human\\User};

\node[
  draw=black, rounded corners=2pt,
  fill=purple!65, fill opacity=0.85,
  text=white, text opacity=1,
  align=center, font=\scriptsize,
  inner sep=1.6pt, text width=1.05cm
] (AI) at (\Gright-0.70,\HdrBoxY) {AI};

\draw[->, line width=1.0pt] (HU.east) -- (AI.west)
  node[midway, fill=white, inner sep=1.2pt, font=\small] {Shift of control};

\tikzset{
  wfbox/.style={
    draw=black,
    fill=white,
    line width=0.45pt,
    inner sep=0.8pt,
    align=center,
    font=\small,
    minimum width=\WboxW
  }
}

\node[wfbox] (S1) at (\Wcx,\Rone)   {\rotatebox{90}{\shortstack{Model\\specification}}};
\node[wfbox] (S2) at (\Wcx,\Rtwo)   {\rotatebox{90}{\shortstack{Input\\modeling}}};
\node[wfbox] (S3) at (\Wcx,\Rthree) {\rotatebox{90}{\shortstack{Implemen-\\tation}}};
\node[wfbox] (S4) at (\Wcx,\Rfour)  {\rotatebox{90}{\shortstack{Simulation\\execution}}};
\node[wfbox] (S5) at (\Wcx,\Rfive)  {\rotatebox{90}{\shortstack{Conducting\\experiments}}};
\node[wfbox] (S6) at (\Wcx,\Rsix)   {\rotatebox{90}{\shortstack{Verification \&\\validation}}};
\node[wfbox] (S7) at (\Wcx,\Rseven) {\rotatebox{90}{\shortstack{Result\\analysis}}};

\draw[->, line width=0.85pt] (S1.south) -- (S2.north);
\draw[->, line width=0.85pt] (S2.south) -- (S3.north);
\draw[->, line width=0.85pt] (S3.south) -- (S4.north);
\draw[->, line width=0.85pt] (S4.south) -- (S5.north);
\draw[->, line width=0.85pt] (S5.south) -- (S6.north);
\draw[->, line width=0.85pt] (S6.south) -- (S7.north);


\node[draw=black, rounded corners=2pt, fill=blue!60, fill opacity=0.78,
      text=white, text opacity=1, align=center, font=\scriptsize,
      inner sep=1.2pt, anchor=west]
  at (\xA,\Rone) {Support for problem\\conceptualization\\(interactively)};

\node[draw=black, rounded corners=2pt, fill=purple!62, fill opacity=0.78,
      text=white, text opacity=1, align=center, font=\scriptsize,
      inner sep=1.4pt, text width=3.05cm]
  at (\xR,\Rone) {Translating\\descriptions into\\formal models\\(fully automatic)};

\node[draw=black, rounded corners=2pt, fill=blue!60, fill opacity=0.78,
      text=white, text opacity=1, align=center, font=\scriptsize,
      inner sep=1.2pt, anchor=west]
  at (\xA,\Rtwo) {Support for\\distribution fitting\\(interactively)};

\node[draw=black, rounded corners=2pt, fill=blue!50!purple!50, fill opacity=0.78,
      text=white, text opacity=1, align=center, font=\scriptsize,
      inner sep=1.4pt, text width=3.00cm]
  at (\xM,\Rtwo) {AI integrated in the\\simulation model\\for direct sampling};

\node[draw=black, rounded corners=2pt, fill=blue!60, fill opacity=0.78,
      text=white, text opacity=1, align=center, font=\scriptsize,
      inner sep=1.2pt, anchor=west]
  at (\xA,\Rthree) {Support for code\\generation\\(interactive)};

\node[draw=black, rounded corners=2pt, fill=purple!62, fill opacity=0.78,
      text=white, text opacity=1, align=center, font=\scriptsize,
      inner sep=1.4pt, text width=2.55cm]
  at (\xR,\Rthree) {Fully automatic\\model generation};

\node[draw=black, rounded corners=2pt, fill=blue!60, fill opacity=0.78,
      text=white, text opacity=1, align=center, font=\scriptsize,
      inner sep=1.4pt, text width=2.70cm]
  at (\xA+2.10,\Rfour) {3D-Scene and asset\\generation};

\node[draw=black, rounded corners=2pt, fill=blue!45!purple!55, fill opacity=0.78,
      text=white, text opacity=1, align=center, font=\scriptsize,
      inner sep=1.4pt, text width=3.05cm]
  at (\xM+0.55,\Rfour) {AI integrated in the\\simulation to\\control model\\behavior};

\node[draw=black, rounded corners=2pt, fill=blue!60, fill opacity=0.78,
      text=white, text opacity=1, align=center, font=\scriptsize,
      inner sep=1.4pt, text width=2.60cm]
  at (\xA+2.05,\Rfive) {Generation of\\experiment plans};

\node[draw=black, rounded corners=2pt, fill=blue!45!purple!55, fill opacity=0.78,
      text=white, text opacity=1, align=center, font=\scriptsize,
      inner sep=1.4pt, text width=3.10cm]
  at (\xM,\Rfive) {Automatic selection\\and execution of\\suitable\\experiments\\(interactively)};

\node[draw=black, rounded corners=2pt, fill=purple!62, fill opacity=0.78,
      text=white, text opacity=1, align=center, font=\scriptsize,
      inner sep=1.4pt, text width=2.70cm]
  at (\xR,\Rfive) {AI approximates\\simulation model\\entirely};

\node[draw=black, rounded corners=2pt, fill=blue!60, fill opacity=0.78,
      text=white, text opacity=1, align=center, font=\scriptsize,
      inner sep=1.2pt, anchor=west]
  at (\xA,\Rsix) {Support for code\\debugging and\\troubleshooting};

\node[draw=black, rounded corners=2pt, fill=blue!45!purple!55, fill opacity=0.78,
      text=white, text opacity=1, align=center, font=\scriptsize,
      inner sep=1.4pt, text width=3.15cm]
  at (\xM+0.65,\Rsix) {Automated analysis\\of result data for\\inconsistent\\behavior for\\validation};

\node[draw=black, rounded corners=2pt, fill=blue!60, fill opacity=0.78,
      text=white, text opacity=1, align=center, font=\scriptsize,
      inner sep=1.4pt, text width=3.00cm]
  at (\xA+2.35,\Rseven) {Extending the users\\toolbox of analysis\\methods};

\node[draw=black, rounded corners=2pt, fill=purple!62, fill opacity=0.78,
      text=white, text opacity=1, align=center, font=\scriptsize,
      inner sep=1.4pt, text width=3.00cm]
  at (\xR,\Rseven) {Fully automatic\\analysis and report\\generation};

\end{tikzpicture}
\caption{Roles that AI can play across the M\&S stages (defined in figure~\ref{fig:sim-study-pipeline}, arranged along a continuum from user-controlled assistance to full AI delegation.}
\label{fig:table2-as-tikz}
\end{figure}

The ways in which AI supports M\&S can be organized based on two orthogonal dimensions. First, AI can be used at different \textit{stages} of the M\&S process (Figure~\ref{fig:sim-study-pipeline}), as discussed previously by Tolk~\cite{tolk2024hybrid}. AI techniques are not used uniformly across stages of the modeling and simulation process, since specific tasks have different requirements such as data needs. For example, LLMs are primarily about text processing, so it is logical that they are used most at stages that are text heavy: building a conceptual model from a text corpus, or transforming data (e.g., model structure, simulation outputs) into text (e.g., executive reports). Conversely, multi-modal LLMs can operate over both text and images, but not every M\&S stage is rich in images so this use has so far been limited to generating images in a report to supplement text~\cite{gandee2024visual} or teaching LLMs about simulation via text and images to support reasoning tasks~\cite{flandre2024can}. Second, AI can play various \textit{roles} within each of the M\&S stages (Figure~\ref{fig:table2-as-tikz}): as an assistant to the modeler, as a necessary piece of the simulation model, or by replacing the simulation model (or the modeler) almost completely. We view these roles as a continuum based on the \textit{extent} to which controls shifts from modelers to AI.

\subsection{Model specification}
Abstracting a system consists of creating a simplified representation that sets clear expectations on objectives, inputs, and outputs, while conveying core assumptions. Deciding what to include (and thus what to leave out) and specifying it through a structure that sets the foundations for a study is a difficult task~\cite{robinson2013conceptual}. This abstraction process is very knowledge-intensive and can involve interdisciplinary teams where subject matter experts discuss key mechanisms and assumptions, which modelers can help to transform into a structured specification. Alternatively, modelers may be exploring the literature to learn about these mechanisms.\footnote{We focus on the most common situation in which a vast amount of data (e.g., interview transcripts or workshop notes when directly accessing experts, articles when using their expertise indirectly) is transformed into a specification by modelers. But that is not the only situation that leads to a conceptual model. For example, a conceptual model can be (re)created from the code using AI to verify the implementation, in a form of reverse engineering~\cite{neykova2025reversed}.} AI can provide valuable support for this time-consuming task. Research has demonstrated the feasibility of using AI (specifically LLMs) to translate a textual description into a variety of models. This use of LLMs as translators can democratize model development by making it accessible to individuals with less modeling expertise~\cite{spillias2025future}. In the case of Agent-Based Models, Onggo \textit{et al.} point out that LLMs make it significantly easier and more intuitive for domain experts to translate complex behavior described in natural language into implementable rules. In this respect, the LLM can perform the transformation work from high-level concept to low-level implementation, which increases the accessibility of ABM to a broader audience~\cite{Onggo2025AIEmpoweredABMS}.

For System Dynamics (SD) and Causal Loop Diagram (CLD), Botello \textit{et al.} indicate that LLMs are fundamentally capable of identifying feedback loops and modeling archetypes from natural language descriptions, but improving performances may call for fine-tuning LLMs for this specific application and investigating the specification of input prompts~\cite{Botello2025AutomatingSDM}. The comprehensive assessment by Schoenberg and colleagues further demonstrates this potential through several benchmarks~\cite{schoenberg2025well}. As the ability to derive models from text becomes more mainstream (e.g., through well-known tools such as {\ttfamily Stella}~\cite{schoenberg2025building}), they are starting to be used for pedagogical purpose~\cite{Fisher2025LLMsSystemDynamicsEducation} and gradually move from an AI assistant to a tutor or co-designer. For example, Stella allows users to `talk' about the model (using {\ttfamily Seldon}), which indirectly prompts LLMs to explain why a model behaves the way it does. The validation and integration of AI for model specification onto later stages of the modeling and simulation process tend to be fragmented across studies~\cite{Muthiah2025AIforSystemDynamics}, which is expected from early works prior to a standardization of practices.

The translation process from text to model does not only reflect the \textit{structure} of the model but also its expected use. For instance, an individual-level model like agent-based modeling needs to know about individuals' attributes, their interactions with each other and their environment; in contrast, an aggregate-level model like system dynamics is concerned about the relations between factors. Yet, two aggregate-level modeling paradigms such as system dynamics and BPMN (Business Process Model and Notation) result in strikingly different AI approaches to translation. Text-to-BPMN approaches prioritize syntactic correctness, unambiguous control flow, and executability, relying on formal intermediate representations such as ASTs or constrained JSON schemas to eliminate ambiguity and ensure standard compliance~\cite{nivon2024automated,licardo2025bpmn}. While BPMN prescribes how processes must execute (e.g., administratively compliant models from legal texts~\cite{macri2025admpmodeler}), CLDs and SD models explain how systems might behave. Thus, text-to-CLD and system dynamics approaches emphasize semantic extraction and conceptual understanding, without necessarily resolving uncertainty in interpretation. This difference also affects the choice of metrics used during evaluation, as a strict adherence to structure and word choices allows for text and graph similarity metrics (e.g., BLEU and ROUGE, Graph Edit Distance~\cite{ccelikmasat2025generating}) whereas CLD/SD admits a multiplicity of valid models and involves benchmarks that consider coverage of key variables and feedback loops, plausibility of causal mechanisms, or qualitative comparisons with reference models.

When employing AI as translators, we leverage their ability to process text and deal with semantic variability, but we do not use their knowledge. The mechanisms and assumptions specified in the conceptual model can be traced back to the knowledge provided as input, such as a corpus of articles. There are also emerging applications in which AI (particularly LLMs) is used as the knowledge source, for instance by acting as experts~\cite{barat2025constructing}. When a conceptual model is learned from data, using LLMs as experts steers the conceptual model towards more selective mechanisms instead of only relying on statistical analyses~\cite{schuerkamp2025guiding}.

\subsection{Input modeling}
After creating specifications, conceptual models, and the like, a simulation study becomes concerned with capturing the input data. This step can be very challenging, especially for stochastic models with complex distribution functions. For example, complex, possibly even irrational behavior of humans or animals can be very difficult to specify (e.g., Resnicow and Vaughn viewed changes in behavior through the lens of chaos theory~\cite{resnicow2006chaotic}). Data-driven AI algorithms are ideal here, as they can approximate these complex distribution functions, provided that sufficient real-world data is available. In principle, there are two paths that can be pursued here: either AI supports the process of input modelling, or it directly replaces a traditional fitted distribution function such that the simulation can sample from the AI.\footnote{In classical stochastic simulation, random inputs such as arrival times, service durations, or routing decisions must be represented by statistical distributions in order to generate new sample paths during simulation runs. These statistical distributions are typically obtained by fitting parametric or empirical models (e.g., Poisson or exponential processes for arrivals) to observed data~\cite{Robinson}[pp.127--133]. Using such models serves both the task of generating values and provides a lever for what-if analyses in varying parameters, for example to explore how a simulation would react if patients in a hospital arrived at more diverse times or the average treatment duration was higher. However, if the primary objective is to accurately reproduce an empirical input rather than support interpretability or hypothesis testing via parameter manipulation, then fitting statistical distributions can be traded for an empirical approach~\cite{Robinson}[pp.134--136]. In such cases, data-driven AI models can be used as input models that learn the underlying distribution directly from data and serve as sampling mechanisms, while acknowledging that this is a trade-off for lower transparency and control in exchange of higher accuracy~\cite{liu2020enhancing}.}

First, AI can support the identification and tuning of suitable distribution functions. This begins with the provision and preparation of the data. Today, large amounts of potentially relevant data for modeling simulation input data are generated from external systems, such as sensors or log files. This data must first undergo appropriate \textit{data cleaning} processes. LLMs have been used to clean data, detect outliers, identify missing values, and even apply and visualize descriptive statistics and metrics~\cite{jansen2025leveraging,zhang2025data}. Another approach by Li and Ji shows how deep learning can be used to directly predict the parameters for a distribution function, which can then be used for input modeling. Specifically, in the context of modeling and simulating construction processes, they trained a neural network to predict the mean and variance of data sets that are merged from heterogeneous sources, such as weather data, machine data, and other construction site conditions~\cite{li2019enhanced}.

The second option for using AI to model input data generates the input data directly, whether for individual samples or even entire time sequences or scenarios. Obviously, generative models are particularly suitable for this purpose. Cen \textit{et al.} for example, present the concept of NIM (Neural Input Modeling), in which variational autoencoders are used in combination with recurrent networks or LSTMs to generate stochastic input data directly. Stochastic processes can be learned directly from historical data. This is particularly advantageous in the case of very complex stochastic relationships that are difficult to map using traditional distribution functions, such as autocorrelations or non-stationary distributions.\footnote{It is relatively rare for a simulation to have an input consisting of either a single value, or a set of fully independent values. Rather, a common situation is to have several values, some of which are related. For example, an agent in an obesity model may have an age, sex, height and weight. There is a general correlation between height and weight, while age and sex are significant predictors of height. Ignoring these relationships would generate agents whose features may be plausible one at a time, but whose \textit{joint} feature values would be unlikely. Manually accounting for all dependencies can be difficult (c.f. Figure 1 in~\cite{beerman2022scoping} as an example from COVID) and fitting a multivariate statistical distribution can be very complex. For instance, we may be fortunate if the correlated attributes all come from a normal distribution and just have different means and/or standard deviations: a \textit{multivariate} normal distribution can then be used. But if each value comes from a different distribution, then finding and fitting a multivariate distribution that captures each aspect becomes its own challenge. For instance, the Johnson distribution is versatile enough to cover normal or lognormal distributions~\cite{debrota1989modeling}, but fitting it needs dedicated methods~\cite{munoz2025comparison}.} In addition, a positive side effect is that the usual requirement for expert knowledge and manual modeling of stochastic distributions in input modeling is reduced$^7$~\cite{cen2020nim,cen2023nim}. Roy \textit{et al.} use GANs to generate trajectories for animal movements during foraging, which can then be used for simulation modeling. The authors report that the GAN was able to reproduce the specific characteristics of seabird foraging trajectories on medium to large spatial scales very well. The quality of the generated trajectories exceeded that of traditional modeling approaches, in particular hidden Markov models. They also point out that, since the GAN training process does not require an explicit probability density of the data, GANs offer a likelihood-free alternative form of modeling for simulation input data and therefore offer new research and application possibilities for modeling those complex input distributions such as animal movements~\cite{roy2022using}. Montevechi \textit{et al.} also use of GANs to generate samples for simulation input data. They confirmed the assumption that GANs work very well for this purpose, especially for approximating complex distributions that are difficult or impossible to model with traditional distribution functions, like highly correlated or multivariate data~\cite{montevechi2021input}. These are just a few examples that show how AI and machine learning can help in modeling input data. Since both the process of input modeling and the training of AI methods are highly data-driven, the two worlds naturally fit together very well here.

\subsection{Implementation of the simulation model}
\label{sec:implementationIntro}
There is also great potential for support in the implementation phase of simulation models. Like in the input modeling, AI methods can be used in several ways: to support the model implementation, as an integrated component during simulation (i.e., AI becomes part of the simulation model and partly controls its output at runtime), or to fully automate the modeling process.

In the first variant, we are mainly talking about support for coding or automatic code generation. Many authors point out that code generation for simulation models using LLMs is fundamentally possible and offers great potential for the future~\cite{du2023ai,frydenlund2024modeler,jackson2024natural}. This use is not limited to supporting modelers in implementing their own model. As pointed out by Monks and colleagues, many models are developed but the code is not always available, so LLMs are a gateway to \textit{model reuse} by re-creating an implementation based on the specifications that another team provided in their article~\cite{monks2025unlocking}.

There are also some interesting approaches and examples for the direct implementation of AI in the simulation model. For example, Woerrlein and Strassburger show an approach in which they train recurrent networks to translate the code for jobs executed by computer numerical control (CNC) machines into power consumption curves. This prediction model can then be integrated into the simulation model of a production line to predict sequences for the power consumption of a production line at runtime in addition to the usual simulation output~\cite{woerrlein2020method}. This is advantageous for a discrete event simulation: power consumption curves are not discrete and they need to be captured precisely to optimize energy consumption, but coupling a discrete event simulation and continuous simulation can be challenging. Training a neural network to translate CNC jobs into power consumption curves provides the required precision while keeping the integration straightforward, as the neural network only has to take the jobs that are running within the simulation and forecast their power consumption curves.

Various papers discuss the idea of using Neural Radiance Fields (NeRFs) to generate three-dimensional animations or scenes in addition to the existing visualization of the simulator.\footnote{Generating \textit{assets} that would be used during a simulation (e.g., 3D elements) would pertain to input modeling as it provides content that goes into a simulation. But generating the \textit{visualization} is a different matter as it pertains to the execution of a simulation process, thus we categorized it under implementation.} This is used intensively in the field of traffic simulation. NeRFs can be used to create 3D assets, including fully-fledged driving environments with high visual quality~\cite{chen2025s}.

\subsection{Simulation execution}
\label{sec:SimulationExecution}
AI and machine learning algorithms can be integrated even more deeply into the model, in such a way that they become an integral part of the model's process logic and, accordingly, the model would no longer be executable without access to the AI. An illustration is provided by Bergmann \textit{et al.} in the context of automatic model generation for approximating dispatching and control rules in manufacturing systems. In order to generate a simulation model automatically, knowledge about the model behavior for priority rules on a sorting buffer does not have to be explicitly modeled, but can be approximated from existing real world log-data using machine learning. To do this, the trained algorithm must be integrated directly into the flow of the simulation model in order to be able to predict appropriate priority decisions at runtime~\cite{bergmann2015approximation}. 

Agent-based modeling and simulation is another field in which it makes sense to implant AI directly into the logic of the model, or more precisely, directly into the agents. This allows complex, more human-like behaviors to be implemented in the agents without having to explicitly codify the process logic and internal rules. Instead, agent behavior is trained and approximated using AI based on real, available data~\cite{dehghanpour2016agent,negahban2021hybrid,platas2023agent}. Generative AI is also predestined for this type of application and is gradually finding its way into agent-based modeling. The term `generative agents' has even been coined for this purpose~\cite{ghaffarzadegan2024generative,park2023generative}. LLMs are particularly helpful here. For example, Ferraro \textit{et al.} show that the integration of LLMs directly into the agent-based model allows for the implementation of realistic social media simulations. In this case, the LLM not only controls agent behavior, but can even retrieve context-dependent external information through retrieval-augmented design~\cite{ferraro2024agent}. Park \textit{et al.} also use an LLM-based architecture so that the AI can control the social behavior of agents. Furthermore, a memory mechanism allows for coherent behavior over a longer period of time, enabling realistic human behavior patterns to emerge, such as conversations, relationships, and group behavior~\cite{park2023generative}. Vezhnevets \textit{et al.} pursue a similar goal and even present a corresponding software library for constructing such agent-based systems~\cite{vezhnevets2023generative}. Recent results suggest that the choice of LLMs and the prompt are very important, as they can generate agents that no longer resemble their human counterparts, or who express a plausible behavior at the aggregate level (e.g., the average simulated response to a scenario is valid) while being incorrect at the individual level (e.g., agents flip between opposite decisions every other day in order for the population average to be maintained)~\cite{11273226}.

There are already initial approaches to the fully automated generation of simulation models using AI and machine learning, albeit so far prototypical or for specific use cases~\cite{behrendt2025real,kute2025generative}. However, these approaches undoubtedly have a great deal of potential and pave the way towards a user-friendly, automated end-to-end framework for carrying out simulation studies, as proposed by Giabbanelli and colleagues~\cite{giabbanelli2024broadening}.

Finally, the AI model can also completely replace the simulation model. This is usually the case in metamodeling~\cite{barton2009simulation}: here, an approximation of the simulation model is trained by a surrogate model using a few (as few as possible) input and output data from the simulation model. This surrogate model can then directly predict the results of further experiments without needing to run the simulation model. This is always useful when the original simulation model is very computationally intensive or has very long runtimes. Executing a metamodel instead of the simulation model can save time and computing resources, as long as it outweighs the efforts required to train the surrogate model. There are numerous examples of using AI for metamodeling. Many modern regression-based machine learning algorithms, including artificial neural networks, have proven effective for this use case~\cite{de2017metamodeling}. Modern generative AI methods can also be used. For example, Cen and Haas use a combination of variational autoencoders and LSTM to generate even complete sequence data as simulation output~\cite{cen2022enhanced}. Transformer models have been used effectively as surrogates for physical simulations, for example fluid mechanics simulations~\cite{geneva2022transformers}.

\subsection{Conducting simulation experiments}
A simulation model has a set of parameters. A \textit{what-if} scenario provides a combination of values for all parameters. For example, a simulation model for the expected benefits from selling a textbook may take as input parameters the expected number of initial buyers (`early adopters'), the adoption rate over time among the broader audience, the unit sale price, production and marketing costs, and the amount of time during which the book may remain relevant. A \textit{what-if scenario provides a specific value to each of these parameters}, thus defining a concrete market situation to simulate expected revenues. Since the future is uncertain, studies tend to consider several what-if scenarios to account for different plausible futures. These scenarios are typically constructed manually, based on domain expertise, stakeholder expectations, or narratives about how key drivers of the system may evolve. As such, what-if analysis is well suited for exploring a small number of meaningful, interpretable futures, but it does not aim to systematically explore the full space of possible parameter combinations. In contrast, \textit{experimental design} approaches treat the parameters as variables spanning a multidimensional search space and generate sets of parameter combinations according to a prescribed design strategy). Rather than representing plausible futures, these combinations are chosen to efficiently probe the model’s behavior, quantify sensitivities, identify interactions among parameters, and support tasks such as optimization, robustness analysis, or model understanding. The field of simulation experiments covers both the creation of a few hand-crafted what-if scenarios and the design of many experiments.

Constructing what-if scenarios is labor-intensive, requires sustained involvement of subject-matter experts, and does not scale well to large or rapidly evolving evidence bases. Text corpora (e.g., academic articles, reports, news) already contain rich, implicit knowledge about drivers, trends, and causal relationships. In the same way as AI uses such sources to create conceptual model, AI can also process text to generate scenarios.\footnote{The design of scenarios is not a single step that can entirely be replaced by AI. There are several phases, and the potential involvement of AI vs. modelers or subject-domain experts varies across phases~\cite{bessa2025integrating}. For example, someone may have to point the AI towards sources that are considered authoritative in an application field, such that it can crawl and build a corpus. Using AI to generate candidate scenarios can be very time-saving, but the proposed scenarios should still be checked by subject-matter experts.} Such text processing abilities became available much before public-facing LLMs such as GPT~\cite{davis2023towards}. Historically, the use of AI to automate the design of scenarios heavily relied on web scraping and topic modeling~\cite{kim2016futuristic,park2016future,kwon2017applying,kayser2020scenario} then it transitioned to artificial neural networks such as {\ttfamily BERT}~\cite{kodding2023scenario,davis2023towards}. Such models can be used through question and answering systems to `interrogate' a text corpus as a means to gradually extract and structure information, thus supporting a scalable and evidence-grounded generation of what-if scenarios~\cite{davis2022automatically,feblowitz2021ibm}. 

Naturally, LLMs are starting to emerge in the space of generating scenarios~\cite{11169695}. In the context of traffic simulation, Ding \textit{et al.} present a system for generating traffic scenarios that can be used for simulators to train autonomous vehicles. This system is a combination of several algorithms that are used to create an AI-based framework for generating traffic scenarios. These generated scenarios include traffic situations, trajectories of agents in the traffic network as well as interaction patterns, such as overtaking maneuvers, turning maneuvers or crash or collision situations~\cite{ding2024realgen}. This work takes a different approach to scenario generation: instead of extracting scenarios from text, Ding and colleagues explored how AI can compose, edit, and control simulation scenarios directly using retrieval-augmented generation in LLMs. In other words, they do not translate scenarios: rather, they \textit{synthesize} scenarios from multiple examples. Noting that scenario \textit{generation} is just one piece of the puzzle, Sadrnezhaad and colleagues discuss the use of generative AI in the context of evaluating, selecting, and integrative scenarios for testing and performing quality assurance of large-scale cyber-physical systems (CPS).\footnote{These are networks of connected systems with close interaction between hardware and software, such as modern vehicles and airplanes. These systems require intensive quality assurance, including testing with simulation~\cite{sadrnezhaad2025generative}.} An important point is that even if AI finds many plausible scenarios, we may not be able to simulate all of them so scenario generation should also be combined with automated means to assess novelty, realism, or relevance (or mechanisms for prioritization and validation)~\cite{sadrnezhaad2025generative}.

Typically, modelers (during development) then decision-makers (during deployment) would be interacting with a simulation model by providing what-if scenarios. They would know what are the parameters and which values can be entered either because they created the model or through their domain expertise. However, other individuals could be affected by decisions derived from a model, so they may want to know what simulations say about their jobs or communities while lacking the technical or domain expertise to directly enter what-if scenarios and run simulations. A broader set of stakeholders\footnote{Stakeholders consist of people who are responsible for, or affected by, the decisions made based on evidence from sources including simulation outputs. There are many other terms such as knowledge users, interested parties, or interest-holders~\cite{akl2024interest}.} could interact with simulations by using LLMs as translators. Users could formulate their requests in plain language (e.g., ``would faculty write more textbooks if people read them?'') and the LLM is tasked with identifying parameter values as well as changes, so that it can instantiate a simulation accordingly~\cite{giabbanelli2024broadening}. 

While what-if scenarios focus on evaluating a small number of plausible futures, the use of AI for experimental deigns aims to explore large sets of parameter combinations. In this setting, the interaction between AI and the simulation model can be structured as a closed loop, where the output of one becomes the input of the other. For example, Feldkamp \textit{et al.} proposed a robustness optimization approach in which GANs are used to automatically generate experimental designs. Robustness optimization attempts to find the best possible settings for control factors so that a target output fluctuates as little as possible and exhibits little variance in response to influences from (in the real world) uncontrollable noise factors. In this concept, two independent GANs are used, which have been trained to generate experimental designs with specific properties. One GAN generates experimental designs for control factors with the aim of increasing the robustness of the system as much as possible, while the other generates experimental designs for the noise factors in order to specifically reduce robustness. The GANs generate experiment plans alternately, and in between, these are executed by the simulation model and the robustness is calculated. After several rounds, an optimal or at least very good, robust solution can be found~\cite{feldkamp2022method}.

A `surrogate' in the context of a simulation study (see Section~\ref{sec:SimulationExecution}) usually means a computationally cheaper AI proxy to an expensive simulation. However, in the context of conducting experiments, the simulation models may be the surrogates because they are cheaper than performing some real-world experiments, particularly in physics. For example, \citet{xu2024automl} explicitly frame simulation as a low-cost experimental surrogate that serves to evaluate and refine experimental designs before applying them to real-world systems. In their work, simulation models are repeatedly executed to generate data under different AI-driven design strategies, allowing robustness, uncertainty, and learning efficiency to be assessed in a controlled setting. So a simulation is not only an object of experimentation, but also at times the enabling platform to design and test AI-driven experimental workflows.

Digital twins provide strong examples of advanced simulators. Digital twins are real-time computational reflections of real, physical systems that (i) continuously collect data from the physical system and (ii) are able to control the physical system~\cite{kritzinger2018digital}. This bi-directional coupling allows for advanced operational intelligence, e.g., using real-time data streams for online analytics and faster-than real-time simulations. Executing simulation experiments with hard real-time constraints is a challenge that requires drawing on AI techniques.

\subsection{Verification and Validation}
Echoing the previous section, AI can again be used in a supporting role or directly integrated into the V\&V process. When it comes to supporting the user, LLMs in particular can once again play to their strengths by lowering entry barriers, as in almost all technical disciplines, by providing expert knowledge quickly and easily. This is particularly useful in verifying implementations, that is, ensuring that the code is correct with respect to specifications. LLMs can be used here, for example, through code generation (see Section~\ref{sec:implementationIntro}), code debugging, and support for systematic troubleshooting~\cite{akhavan2024generative,du2023ai}. In the same manner as AI companions (e.g., Microsoft Copilot) can provide assistance in general programming environment, LLMs can be adapted to simulation environments as illustrated by {\ttfamily NetLogo Chat}~\cite{chen2024learning}.\footnote{Whether the general use of AI companions helps to break barriers or creates dependency is a broad debate, much beyond the scope of modeling and simulation~\cite{zhang2025breaking}. On the one hand, we can argue that novice modelers can now more easily create models and learn from simulation experiments, which benefits them because early engagement with runnable models and data-driven experimentation can strengthen conceptual understanding and motivation. On the other hand, there is an erosion of expertise as hands-on experiences are replaced by a reliance on AI companions to write simulation code~\cite{spillias2025future}. This may also affect quality, as novices may not learn and ensure whether the code is \textit{effective} (does the right things) or \textit{efficient} (uses minimal resources). The comparison of novices and experts using the AI companion {\ttfamily NetLogo Chat} exemplifies these concerns: experts \textit{selectively} copy code produced by the LLM and debug it at least partly \textit{themselves}, while novices tend to copy the whole code and debug it with the help of AI~\cite{chen2024learning}. Eventually, these concerns call for changes in how we teach modeling and simulation with AI. As with the earlier integration of big data into M\&S education, the key challenge is not whether AI should be used, but how responsibilities are distributed between the learner and the tool: AI can handle syntactic complexity, while curricula must still explicitly train students to reason about model structure, assumptions, validity, and computational trade-offs~\cite{giabbanelli2016teaching}.}

In the validation phase, we check the correctness of the data generated by the simulation model, for example by comparing it with any real data that may be available. AI can also help here, for example, by using algorithms to detect anomalies and outliers. In a sufficiently large amount of simulation data, unexpected or inconsistent model behavior can be found relatively quickly~\cite{feldkamp2020knowledge}. He \textit{et al.} show that LLMs can help extract the right latent variables from real-world data in order to better and more validly map the behavior of agents in agent-based modeling, strengthening the validity of the model~\cite{He2025LLMModelValidation}.

AI can also help sift through patterns that provide a more fine-grained evaluation but would otherwise be ignored by typical analyses. For example, we commonly evaluate simulation outputs at the aggregate level: in a forest fire simulation, did the expected number of trees burn (either over time or when the fire was controlled)? While it is important to ensure that aggregate numbers are aligned with expectations, it is not \textit{sufficient} to conclude that a simulation is valid, because incorrect implementations can still produce aggregate outputs that fall within expected distributions. Detailed patterns matter: a forest fire simulation may be correct as an aggregate but have an incorrect understanding of the effect of wind on embers, such that it simulates the fire \textit{going in the wrong direction}. This matters for decision-making, since the point of a fire simulation is not usually to watch trees as they burn, but to \textit{intervene in specific places} to protect lives and resources. When using simulation techniques such as cellular automata, individual cells are updated over time so the simulation produces spatio-temporal data that tracks how the fire spreads at a fine-grained level. While that data is usually aggregated for analyses, AI provides opportunities to use it fully. These opportunities require transforming simulation outputs into formats aligned with AI algorithms. In particular, commonly used cellular automata (consisting of a 2D grid of square cells) can be transformed into images by mapping the state of each cell to a color. This unlocks access to a wide array of image processing algorithms~\cite{wozniak2021comparing}, and even AI analyses to track patterns in a simulation run as if it was a video.\footnote{The goal is not the visualization, but its ability to support downstream tasks such as finding implementation errors that would be invisible at the aggregate level. By transforming simulation traces into images and applying AI techniques such as machine learning classifiers, subtle spatial or topological inconsistencies (e.g., missing interactions, incorrect update orders) can be identified even when summary statistics appear correct.} Using AI to inspect simulation outputs as images is not limited to physical phenomena such as forest fires, flooding, or landslides. Even phenomena that do not have a physical embodiment, such as network-based social simulations, can be transformed into image-like representations that preserve the structure of local interactions for AI-based analysis~\cite{wozniak2022new}.\footnote{Again, the objective is not to `produce an image', but to render simulation outputs in a way that aligns with the input needs for AI algorithms. Visualizing a large-scale network simulation as a node-and-link diagram would miss the point because this would not support algorithms in clearly finding patterns. In the same manner as a cellular automaton maps to an image (cells become pixels), a network can be viewed as a matrix and that becomes an image. However, the same network can be represented through multiple matrices (e.g., rows and columns can be swapped while encoding the same object) so a sub-problem for AI is to determine the right representation of simulation data.}

\subsection{Result Analysis}
During the analysis phase of simulation results, the two levels at which AI can be used become apparent once again, i.e., to support the user in the analysis of the result data or directly integrate AI into the process of output data analysis. In the first case, LLMs are once again a good choice. There is currently still a lot of research potential here. For example, LLMs could automatically summarize results, write reports, or prepare project documentation~\cite{akhavan2024generative,giabbanelli2023gpt}. It is also conceivable that LLMs could advise and support users in statistical analyses and evaluations, similar to the support process for input modeling.

As already mentioned, AI and machine learning methods can also be used as analysis tools and become an integral part of the analysis. For example, Curran \textit{et al.} use machine learning to automatically classify the output of system dynamics models into pre-defined classes of behavioral patterns. This means that even large simulation experiment data can be analyzed efficiently and automatically. Since their approach used a decision tree, a secondary benefit is that the reasons for the classification results are transparent,\footnote{Decision trees are a classic example of a white-box prediction algorithm, because their internal decision logic is explicitly represented as a sequence of hierarchical tests on input variables. Each path from the root of the tree to a leaf corresponds to a concrete decision process that can be followed visually as a diagram or equivalently expressed as a set of human-readable if–then rules. This makes it straightforward for analysts to trace how specific input conditions lead to a particular classification outcome, inspect which variables and thresholds are decisive, and assess whether these decision rules are plausible given domain knowledge.} enabling the generation of additional insights into the behavior of the underlying simulation model~\cite{curran2025classification}. Generating knowledge about the system can therefore be the main reason for applying such algorithms to simulation data, so that the actual prediction function then no longer plays such an important role. This approach requires conducting a sufficient number of simulation experiments and training the prediction algorithm accordingly with simulation input data and associated result data~\cite{feldkamp2020knowledge}. Classification and regression trees are predestined for this purpose~\cite{philippe2010impact}. However, white-box models such as decision trees have limited expressive power and may fail to capture complex, nonlinear relationships that arise in simulation data. Decision trees rely on piecewise linear partitions of the input space, which can obscure interactions or curved decision boundaries. More expressive models, such as support vector machines or neural networks, can better approximate such relationships, but that comes at the cost of interpretability. As a result, analysts must balance the goal of understanding simulation behavior against the goal of accurately modeling it, often resorting to black-box models augmented with explainable AI (XAI) techniques~\cite{feldkamp2022explainable,serre2021use}.

\section{Modeling \& Simulation for AI}

After we discussed the utility of AI for M\&S, we focus on the other direction: how M\&S benefits AI. Combining M\&S with AI has numerous advantages, typically related to the training, validation, and evaluation of AI models.

\subsection{Training AI models on simulated data}
The success of modern, subsymbolic AI hinges on the volume and quality of available data used for training and validation. However, data is often too costly to acquire, which limits the potential of the developed AI system. Training AI models on simulated data offers a cost-efficient and safe alternative to manually collecting data from operational environments. In such a setup, the simulator is used for generating data in an off-line fashion, i.e., decoupled from AI agent under training.

AI models can be trained directly on simulators, too, i.e., in an on-line fashion. In such a setup, the simulator provides an environment or exercisable interface through which AI agents can be trained, tested, and validated before being deployed into a production environment. The utility of off-line and on-line training depends on the specific machine learning approach. Conventional deep learning, for example, requires large volumes of data to be available; such needs are easily served by off-line training. Reinforcement learning, in contrast, requires interactions with the simulator and is better served by on-line training.

\subsection{Training AI models in Digital Twins}

The proliferation of digital twins~\cite{kritzinger2018digital,david2024infonomics} opened new perspectives in using simulations for AI training~\cite{liu2025ai}. In addition to providing operational intelligence services, digital twins may offer services for training AI models on their simulators. The conceptual and architectural organization of digital twins allow for such scenarios naturally. In addition, digital twins are able to control the physical system they are coupled with. This control is traditionally used for the real-time optimization of the system (e.g., for energy-efficiency), but can be also used for bringing the physical system into states that provide new insights into the system's behavior. This allows for gathering data outside the validity domain of the digital twin's simulator. Such a \textit{purposeful experimentation} is not new in modeling\&simulation: the concept first appeared in Zeigler's seminal work on simulation~\cite{zeigler2000theory}, and a new wave of research has been investigating its theoretical foundations recently~\cite{mittal2023towards}. Digital twins gave a renewed momentum to the field by aiding the operationalization of purposeful experimentation and by that, paving the foundations for cyber-physical AI training environments.

Reinforcement learning is an apt example of AI training techniques that require more than mere generation of data. In reinforcement learning~\cite{SuttonBarto2018ReinforcementLearning}, an AI agent learns optimal control of an environment by taking sequential actions to explore the environment and update its strategy based on feedback in the form of rewards. Such a trial-and-error approach is costly and hazardous when executed in a real-world setup, e.g., when an autonomous vehicle learns to drive~\cite{isele2018safe} or a cyber-biophysical system is being actuated~\cite{david2023digital}. Situating this learning process in a digital twin has clear benefits, as explained above. In addition, digital twins themselves can be equipped with learning capabilities that allow for the gradual improvement of their simulators~\cite{david2024automated,david2022devs}, borrowing them useful evolutionary capabilities~\cite{david2023towards}, which is particularly useful in smart ecosystems, such energy communities~\cite{michael2024smart} and the smart grid~\cite{liu2026introduction}.

Model predictive control (MPC) takes a similar approach~\cite{wu2025tutorial}: in systems and processes with constraints (usually industrial processes), predictions are used to repeatedly find and execute optimal control rules within a timely prediction horizon. A simulation model is essential for calculating these predictions. In order to be able to implement even complex controls live and online, AI and machine learning algorithms are increasingly being used here as well, but these in turn rely on training data from the simulation model. The AI does not necessarily have to replace the simulation model, but can also complement it~\cite{lawrynczuk2025lstm,kumar2018deep}. An AI/simulation combination approach can also be used for AI-driven design processes (e.g., in materials science or photonics). In so-called inverse design, an AI algorithm suggests appropriate designs, which are then evaluated and validated by simulation, allowing the AI to iterate again in a data-driven manner~\cite{jin2025machine}. Finally, simulation can also be a necessary safety requirement for AI systems that are integrated into technical or mechatronic systems, for example. Since AI systems are often considered black boxes due to their lack of transparency, they can therefore only be evaluated analytically to a limited extent. For this purpose, Dahmen et al. propose simulation-based test beds that can provide extensive virtual scenarios and data sets to improve the quality assurance of those systems~\cite{dahmen2023structured}.

\subsection{Evaluating AI by Modeling \& Simulation}

Finally, M\&S can be of high utility in the evaluation and quantified assessment of AI. This lifecycle phase of AI systems remains a significant challenge due to the stochastic behavior, data dependence, and limited transparency of AI models. M\&S can play a critical role in tackling these challenges by providing a controlled, repeatable, and scalable way for driving the behavior of AI models for observation and measurement. Through simulation, researchers can expose AI systems to a wide range of scenarios, including rare, extreme, or safety-critical conditions that are difficult or costly to reproduce in the real world.

Beyond the simulation models themselves and their integration with AI, additional aspects of  M\&S offer valuable opportunities for machine learning and AI, which can be leveraged in the development of such algorithms. For example, M\&S offers decades of experience in efficient  experiment design. Sanchez therefore proposes to make use of this knowledge and to employ sophisticated data farming experiment designs to make the hyperparameter tuning of machine learning algorithms such as neural networks more efficient. With a NOLH (Nearly Orthogonal Latin Hypercube) design, the space of possible hyperparameters can be searched extremely efficiently and with a fraction of the effort compared to a complete iteration of all hyperparameter combinations~\cite{sanchez2020data}.

\section{Perspectives: Evolution of the field}
\begin{figure}[htbp]
    \centering
    \includegraphics[width=\textwidth]{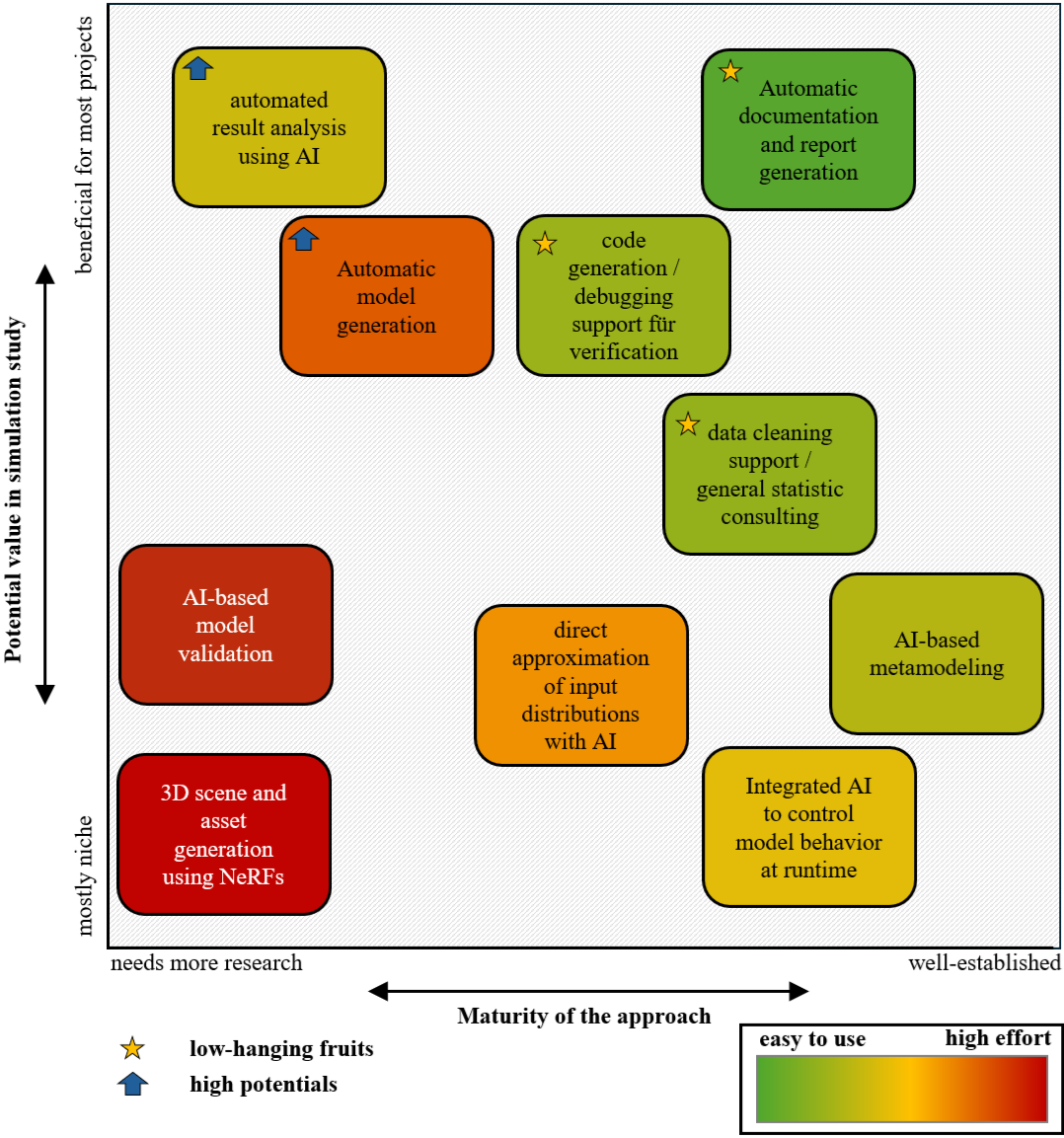}
    \caption{Landscape of artificial intelligence applications in modeling and simulation, organized by maturity of the approach (from exploratory to well-established) and potential value in simulation studies (from niche to broadly beneficial).}
    \label{fig:AIquadrant}
\end{figure}

The preceding sections surveyed how AI has already been used to support, augment, or integrate M\&S across the stages of a simulation study. This overview was not intended to be exhaustive, nor to suggest that these approaches are uniformly mature or widely adopted. Rather, its purpose was to establish a conceptual structure for thinking about the many roles AI can play in M\&S. In this final section, we shift from retrospect to perspective. Instead of asking what has already been achieved, we consider where the field is heading and which AI-enabled approaches appear most promising in terms of practical impact, accessibility, and research potential. Figure~\ref{fig:AIquadrant} provides a high-level positioning of these approaches, contrasting their current maturity with their potential value for simulation studies. While some uses of AI are already well established and readily accessible, others remain exploratory, requiring substantial effort and further research before they can be reliably integrated into everyday modeling and simulation workflows.

Although the previous sections already showed that LLMs are increasingly used, we believe that there is more room for LLMs to help modelers. For instance, consider result documentation and report generation (from existing results). Access to powerful LLMs to handle such tasks is easy, their integration can be done with careful engineering, and the benefits in terms of time savings are obvious. The same applies to LLM-supported code generation in the implementation phase and LLM-supported code debugging in the model verification phase. LLM-supported coding is already widely established and its performance has been confirmed by benchmarks. However, performance here also depends on the programming language or simulation software used. The extent and quality of data and examples available on the internet that could serve as the basis for the language model's training data is obviously relevant for its performance. For widely used programming languages such as Python or Java, good availability can generally be assumed. For less common programming languages, specific simulation packages, or specialized simulators, however, the data available is likely to be limited, thereby reducing the quality of the LLM responses or even leading to hallucinations. The situation is similar with LLM-based support for data cleaning and general support for mathematical or statistical questions. LLMs have already proven helpful for assisting the user in standardized data cleaning tasks~\cite{spreafico2025lost}. Since data cleaning is the start of every data science projects, there is likely a good coverage of use cases in the training data for common LLMs, which would explain their performances. These approaches are thus applicable across simulation studies. Of course, it is essential that users critically examine outputs produced by LLMs. For this reason, caution is needed when aiming for a fully automatic result analysis using LLMs. Although this approach promises the most future potential according to a recent literature study, it also still requires significantly more research and further development at this stage~\cite{Feldkamp2025GenerativeAIinSimulation}. This is because LLMs are currently very well suited for processing and generating language (and code), but tend to struggle when it comes to more complex mathematical tasks and advanced statistical reasoning as current studies and benchmarks show~\cite{liu-etal-2024-llms}. In this respect, LLMs can provide very good conceptual advice and explanations, but they are not yet suitable for fully automated analysis of simulation results.

In contrast, metamodeling is a mature and well-established approach, and using AI in this context has been extensively studied. However, its applicability in a given simulation project should be evaluated primarily through a cost–benefit lens: the additional effort required for model development, training, hyperparameter tuning, and quality assurance must be justified by a corresponding reduction in simulation effort. For complex AI-based metamodels, this overhead may offset or even exceed the savings in simulation run time. Similar considerations apply to approaches that directly approximate input distributions for sample generation. Their usefulness depends on factors such as data availability, the complexity of the target distribution, and the expected gain in accuracy. While AI-based methods can be advantageous for complex, multimodal, or high-dimensional distributions, traditional statistical techniques often remain the more efficient choice in simpler settings.

A similar assessment applies to integrating AI directly into simulation models at runtime. Such integrations are not new and have long been used in specific contexts, but their usefulness remains highly application-dependent. Because design choices and computational demands vary widely, the effort–benefit trade-off must be evaluated for each use case. At the same time, recent advances in generative AI are driving renewed interest in this paradigm, particularly in agent-based modeling. Approaches such as generative agents suggest new ways of equipping simulated entities with adaptive, data-driven behavior, making this an area with significant potential for future research~\cite{Feldkamp2025GenerativeAIinSimulation}.

Fully automatic model generation continues to be a challenging topic, especially when using component-based simulation frameworks and not just pure programming language. Nevertheless, there is a growing trend toward integrating such functions into commercial off-the-shelf (COTS) software, which indicates a growing relevance in both research and practice. Dynamic developments are expected in the near future. 
In contrast, AI approaches for the validation of simulation models are still relatively rare. In addition, the idea of using AI in simulation models raises additional questions about the validity of the resulting simulation results, for which further research is needed. Finaly, the generation of 3D scenes and assets using NeRFs is an exciting approach, but ultimately it should be considered more of a niche topic. In particular, the resulting benefits must be weighed against the very high development and computing costs that typically accompany such algorithms.

Our observations suggest that the influence of AI on modeling and simulation is not merely incremental, but potentially transformative. Over time, the field has repeatedly evolved in response to new methodological foundations, from early concerns with queuing theory and random number generation, to data-driven modeling and machine learning, and now to the rapid emergence of large language models. These shifts challenge established practices and, at times, the very identity of modeling and simulation as a discipline. Rather than signaling a loss of rigor or purpose, however, they reflect an \textit{expanding methodological toolkit} and an opportunity to reconsider what it means to be a simulation scientist in a data- and AI-rich world.

\bibliographystyle{plainnat}
\bibliography{mybib}
\end{document}